\documentclass[twoside]{article}

\usepackage[accepted]{mysty}
\usepackage[round]{natbib}

\usepackage{amsmath, amssymb, amsthm}
\usepackage{algorithm}
\usepackage{algorithmicx}
\usepackage{algpseudocode}
\usepackage{enumerate}
\usepackage{booktabs}
\usepackage{graphicx}
\usepackage{bm}

\newcommand{\Normal}{\operatorname{Normal}}
\newcommand{\Reals}{\mathbb R}
\newcommand{\zeros}{\bm 0}
\newcommand{\Identity}{\mathrm I}
\newcommand{\iid}{\stackrel{\text{iid}}{\sim}}
\newcommand{\indep}{\stackrel{\text{ind}}{\sim}}
\newcommand{\Tree}{\mathcal T}
\newcommand{\sM}{\mathcal M}
\newcommand{\Leaves}{\mathcal L}
\newcommand{\Uniform}{\operatorname{Uniform}}
\newcommand{\Exponential}{\operatorname{Exponential}}

\newcommand{\Dirichlet}{\operatorname{Dirichlet}}
\newcommand{\supp}[1]{^{(#1)}}
\newcommand{\DP}{\operatorname{DP}}
\newcommand{\Categorical}{\operatorname{Categorical}}
\newcommand{\Var}{\operatorname{Var}}

\RequirePackage[pdfusetitle, colorlinks,citecolor=blue,urlcolor=blue]{hyperref}
\begin{document}



\twocolumn[

\aistatstitle{Interaction Detection with Bayesian Decision Tree Ensembles}

\aistatsauthor{ Junliang Du \And Antonio R. Linero}

\aistatsaddress{ Department of Statistics, Florida State University} 

]

\begin{abstract}
  Methods based on Bayesian decision tree ensembles have proven valuable in constructing high-quality predictions, and are particularly attractive in certain settings because they encourage low-order interaction effects. Despite adapting to the presence of low-order interactions for prediction purpose, we show that Bayesian decision tree ensembles are generally anti-conservative for the purpose of conducting interaction detection. We address this problem by introducing Dirichlet process forests (DP-Forests), which leverage the presence of low-order interactions by clustering the trees so that trees within the same cluster focus on detecting a specific interaction. We show on both simulated and benchmark data that DP-Forests perform well relative to existing interaction detection techniques for detecting low-order interactions, attaining very low false-positive and false-negative rates while maintaining the same performance for prediction using a comparable computational budget.
\end{abstract}

\section{INTRODUCTION}

In many scientific problems, a primary goal is to discover structures which allow the problem to be described parsimoniously. For example, one may wish to find a small subset of candidate variables that are predictive of a response of interest; this structure is referred to as \emph{sparsity}. Another structure is \emph{interaction} (or \emph{additive}) structure. An extreme case of additive structure is a generalized additive model (see, e.g., \citealp{hastie2017generalized}), where the effects of the predictors combine additively without any interactions. Teasing out additive structures can be valuable because it can substantially simplify the interpretation of a model. For example, if a given predictor does not interact with other predictors then it can be interpreted in isolation without reference to the values of other predictors. When predictors do interact, interpretation of the interactions is typically simplified whenever the interactions are of low-order. We consider the nonparametric regression problem $Y_i = f_0(X_i) + \epsilon_i$, $\epsilon_i \sim \Normal(0,\sigma^2)$,
% \begin{align*}
%     Y_i = f_0(X_i) + \epsilon_i, \qquad \epsilon_i \sim \Normal(0, \sigma^2), 
% \end{align*}
where $Y_i$ is a response of interest and $X_i \in \Reals^P$ is a vector of predictors, however the methods we develop here can be easily extended to many other settings. The variables $x_j$ and $x_k$ are said to \emph{interact} if $f_0(x)$ cannot be written as $f_0(x) = f_{0\backslash j}(x) + f_{0\backslash k}(x)$ where $f_{0\backslash j}$ and $f_{0\backslash k}$ do not depend on $x_j$ and $x_k$ respectively. One can define higher order interactions similarly: a group of $K$ variables is said to have a $K$-way interaction if $f_0(x)$ cannot be decomposed as a sum of $K$ or fewer functions, each of which depends on fewer than $K$ of the variables. 

Methods which estimate $f_0(x)$ using an ensemble of Bayesian decision trees have proven useful in a number of statistical problems. Beginning with the seminal work of \citet{chipman2010bart}, Bayesian additive regression trees (BART) have been successfully applied in a diverse range of settings including survival analysis \citep{sparapani2016nonparametric}, causal inference \citep{hahn2017bayesian}, variable selection in high dimensional settings \citep{linero2016bayesian, bleich2014variable}, loglinear models \citep{murray2017log}, and analysis of functional data \citep{starling2018functional}. A key motivating factor for the use of BART is precisely that it is designed to taking advantage of low-order interactions in the data. Indeed, \citet{linero2017abayesian} and \citet{rockova2017posterior} illustrate theoretically that the presence of low-order interactions is precisely the type of structure which BART excels at capturing. Hence BART appears to be an ideal tool for extracting low-order and potentially non-linear interactions.

Surprisingly, we show that, despite the ability of BART to capture low-order interactions for \emph{prediction} purposes, it is nonetheless not suitable for conducting fully-Bayesian inference for the \emph{selection} task of interaction detection. When taken at face value as a Bayesian model, we show empirically that BART generally leads to the detection of spurious interaction effects. This is not contradictory because optimal prediction accuracy is generally \emph{not} sufficient to guarantee consistency in variable selection (see, e.g., \citealp{wang2007tuning}). 

We discuss the general problem which leads to the detection of spurious interactions; while this development is couched in the BART framework, we believe that the fundamental issues also occur for other decision tree ensembling methods. Specifically, the problem is that there is no penalty associated to including spurious interaction terms in the model.
We then introduce a suitable modification to the BART framework which addresses this problem and allows BART detect interactions in a fully-Bayesian fashion. We accomplish this by clustering the trees into non-overlapping groups. Intuitively, the shallow trees comprising each cluster work together to learn a single low-order interaction. To bypass the need to specify the number of clusters, we induce the clustering through a Dirichlet process prior \citep{ferguson1973}. We refer to the ensemble constructed in this fashion as a Dirichlet Process Forest (DP-Forest).

\subsection{A Simple Example}
\label{sec:a-simple-example}

\begin{figure}[t]
    \centering
    \includegraphics[width=.45\textwidth]{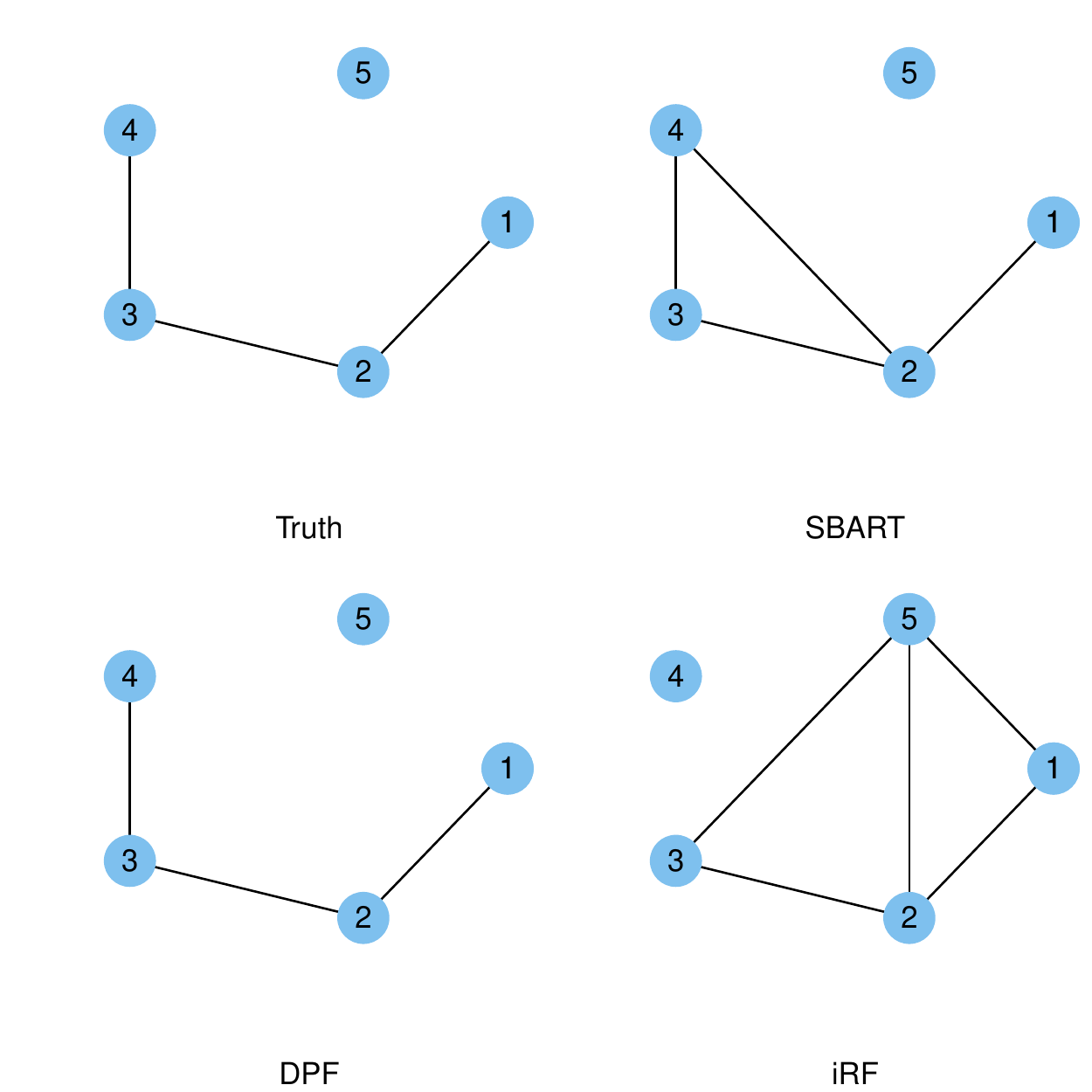}
    \caption{The interaction structure detected in the example from Section~\ref{sec:a-simple-example}. ``Truth'' denotes the true interaction structure in the example.}
    \label{fig:interaction-graph}
\end{figure}

To motivate the problem, we consider a simulated data example of \citet{vo2016sparse}. This example takes $P = 100$, $N = 100$, $X_i \sim \Normal(\zeros, 0.02\,\Identity)$, and 
$f_0(x) = x_1 + x_2^2 + x_3 + x_4^2 + x_5 + x_1x_2 + x_2 x_3 + x_3x_4$.
We compare the DP-Forest we propose to a variant of BART referred to as SBART \citep{linero2017abayesian} which can accommodate sparsity in variable selection. We also consider the recently proposed iterative random forests algorithm of \citet{basu2018iterative}, selecting interactions whose stability score is higher than 0.5. In Figure~\ref{fig:interaction-graph} we display the interaction structure detected by each method on this data; while we considered only one iteration of this experiment here, these results are typical of replications of the experiment. 

Here, SBART detects a spurious edge between $x_2$ and $x_4$. This occurs because BART, despite its fundamentally additive nature, does not include any penalization which discourages unnecessary interactions from being included. On the contrary, BART \emph{expect} interactions to occur between relevant predictions; considering a draw from a BART prior such that $x_2$ and $x_4$ are included in the model, an interaction between these variables is a-priori likely. Adapting Bayesian decision tree ensembles to interaction detection then requires a prior which discourages the inclusion of weak interactions. The iRF similarly detects two spurious interactions and misses a relevant interaction between $x_3$ and $x_4$. 

\subsection{Related Work}
\label{sec:related-work}

Recent work has studied the theoretical properties of BART. \citet{linero2017abayesian} and \citet{rockova2017posterior} show that certain variants of BART are capable of adaptively attaining near-minimax-optimal rates of posterior concentration when $f_0$ can be expressed as a sum of low-order interaction terms $f_0(x) = \sum_{v = 1}^V f_{0v}(x)$ with each $f_{0v}(x)$ depending on a small subset of $\mathcal S_v$ of the predictors.
In view this, one might conclude that no modification to BART is needed. This is true if one cares only about the mean integrated squared error $\int (f_0(x) - f(x))^2 \, F_0(dx)$ where $X_i \iid F_0$. Optimal prediction performance, however, does not imply that variable selection and interaction detection are being performed adequately. If $\mathcal S_0$ is the true interaction structure of the data $\mathcal S$ is an estimate of $\mathcal S_0$, then attaining the minimax estimation rate for $f_0$ in terms of prediction error typically only guarantees that $\mathcal S_0 \subseteq \mathcal S$ (not $\mathcal S \subseteq \mathcal S_0$).

Several other methods have been recently proposed in the literature specifically for the task of interaction detection. We offer a non-comprehensive review. For a recent review, see \citet{bien2013lasso}. \citet{lim2015learning} proposed a hierarchical group-lasso which enforces the constraint that the presence of a given interaction implies the presence of the associated main effects; a similar approach is given by \citet{bien2013lasso}. A potential shortcoming of these approaches is that they focus on linear models and allow only pairwise interactions. \citet{radchenko2010variable} propose the VANISH algorithm, which allows for nonlinear effects through the use of basis function expansions, but again limits to pairwise interactions. Several decision-tree based methods have also been proposed. The additive groves procedure of \citet{sorokina2008detecting} uses an adaptive boosting-type algorithm to sequentially test for the presence of interactions between variables after performing a variable screening step. \citet{basu2018iterative} propose the iterative random forest (iRF) algorithm which flags ``stable'' interaction effects as those which appear consistently in many trees in a certain random forest.

\section{BAYESIAN TREE ENSEMBLES}
\label{sec:review}

\subsection{The BART Prior}

Our starting point is the Bayesian additive regression trees (BART) framework of \citet{chipman2010bart}, which treats the function $f_0(\cdot)$ as the realization of a sum of random decision trees
\begin{align*}
  f(x) = \sum_{t = 1}^T g(x ; \Tree_t, \sM_t),
\end{align*}
where $\Tree_t$ denotes the tree structure (including the decision rules) of the $t^{\text{th}}$ tree and $\sM_t = \{\mu_{t\ell} : \ell \in \Leaves_t\}$ denotes the parameters associated to the leaf nodes; here, $\Leaves_t$ denotes the collection of leaf nodes of $\Tree_t$. Let $[x \leadsto (t,\ell)]$ denote the event that the point $x$ is associated to leaf $\ell$ in tree $t$. The function $g(x; \Tree_t, \sM_t)$ then returns $\mu_{t\ell}$ whenever $[x \leadsto (t,\ell)]$ occurs.% , i.e., $g(x ; \Tree_t, \sM_t) = \sum_{\ell \in \Leaves_t} \mu_{t\ell} I(x \leadsto (t,\ell))$. 

% \begin{figure}
%   \centering
%   \includegraphics[width=.5\textwidth]{figure/GraphFigure}
%   \caption{Schematic showing the generation of a regression tree $(\Tree, \sM)$ from the prior (bottom) with the associated recursive partition of the predictor space $\mathcal X = [0,1]^2$. After the tree topology and splitting rules are generated, a parameter $\mu_\ell$ associated to leaf node $\ell$ is drawn.} 
%   \label{fig:GraphFigure}
% \end{figure}

We follow \citet{chipman2010bart} and specify a branching process prior for the tree structure $\Tree_t$.  
% The terms $(\Tree_1, \sM_t), \ldots, (\Tree_T, \sM_T)$ are independent and identically distributed (conditional on hyperparameters). 
A sample from the prior for $\Tree_t$ is generated iteratively, starting from a tree with a single node of depth $d = 0$; this is made a branch with two children with probability $q(d) = \gamma / (1 + \beta)^d$, and is made a leaf node otherwise. We repeat this process independently for all nodes of depth $d = 1, 2, \ldots$ until all nodes at depth $d$ are leaves. After the structure of the tree is generated, each branch $b$ is associated with a decision rule of the form $[x_j \le C_b]$. The coordinate $j$ used to construct the decision rule is sampled with probability $s_j$ where $s = (s_1, \ldots, s_P)$ is a probability vector. The splitting proportion $s$ will play a key role later as an avenue for inducing sparsity in the regression function. Finally, we generate $C_b \sim \Uniform(L_j, U_j)$ where $(L_1, U_1) \times \cdots \times (L_P, U_P)$ is the hyper-rectangle corresponding to the values of $x$ that lead to branch $b$. We remark that this choice for $C_b$ differs from the scheme used by other BART implementations; we adopt it to simplify the full conditionals we derive in Section~\ref{sec:dirichlet}. 

For the prior on $\sM_t$ we set $\mu_{t\ell} \iid \Normal(0, \sigma^2_\mu / T)$ conditional on $\Tree_t$ and $\sigma^2_\mu$. By taking the variance to be $\sigma_\mu^2/T$ we ensure that the prior level of signal is constant as $T$ increases. The normal prior is selected for its conjugacy; we note, however, that any prior for $\mu_{t\ell}$ with mean $0$ and variance $\sigma_\mu/T$ leads to the approximation $f(x) \sim \Normal(0, \sigma^2_\mu)$ by the central limit theorem. We fix $\beta = 2$ and $\gamma = 0.95$; we refer readers to \citet{linero2017abayesian} for further details regarding prior specification, and to \citet{chipman2013bayesian} and \citet{linero2017review} for detailed reviews of Bayesian decision tree methods.

% A benefit of the BART framework is that there exist default prior specifications which are highly effective in practice. We fix $\beta = 2$ and $\omega = 0.95$; we refer readers to \citet{linero2017abayesian} for further details regarding prior specification, and to \citet{chipman2013bayesian} and \citet{linero2017review} for detailed reviews of Bayesian decision tree methods.

% A benefit of the BART framework is that there exist default prior specifications which are highly effective in practice. We assume that $Y_i$'s have been scaled to have mean $0$ and standard deviation $1$ and that the $X_i$'s have been scaled using a quantile transformation to lie in $[0,1]$. We fix $\beta = 2$ and $\gamma = 0.95$ and use a half-Cauchy prior $\sigma_\mu \sim \Cauchy_+(0, 1)$ for $\sigma_\mu$. We use an empirical prior for the error variance $\sigma$, setting $\sigma \sim \Cauchy_+(0, \widehat \sigma)$ where $\widehat \sigma$ is an empirical estimate of $\sigma$ obtained from, e.g., fitting the lasso. For more detailed reviews of Bayesian decision tree based methods, we refer readers to \citet{chipman2013bayesian} and \citet{linero2017review}.

\subsection{Leveraging Structural Information}

Several recent developments have extended the BART methodology to take advantage of structural information. \citet{linero2016bayesian} noted that sparsity in $f_0(x)$ can be accommodated automatically by setting $s \sim \Dirichlet(\alpha / P, \ldots, \alpha / P)$. Recall here that $s_j$ denotes the prior probability that, for a fixed branch, coordinate $j$ will be used to construct a split at a that branch. Hence, if $s$ is nearly-sparse with $d$ non-sparse entries, the prior will encourage realizations from the prior to include only the $d$ predictors with non-sparse entries. 
% The sparsty-inducing Dirichlet prior encourages this behavior; realizations from the sparsity-inducing prior with $P = 3$ are given in Figure~\ref{fig:sparsity}. Conveniently, under the prior described above for $\Tree_t$, the conjugacy of the Dirichlet prior to multinomial sampling leads to a Dirichlet full conditional for $s$. 
\citet{linero2017abayesian} showed that this prior for $s$ induces highly desirable posterior concentration properties; in particular, the posterior of $f(x)$ concentrates at close to the oracle minimax rate if we had known the relevant predictors beforehand. 

% \begin{figure}
%     \centering
%     \includegraphics[width=.17\textwidth]{./figure/Sparsity1}
%     \includegraphics[width=.17\textwidth]{./figure/Sparsity3}
%     \includegraphics[width=.17\textwidth]{./figure/Sparsity33}
%     \caption{Draws from $\Dirichlet(\alpha/3, \alpha/3, \alpha/3)$ priors for differing values of $\alpha$. Verticies correspond to one-sparse vectors, edges to two-sparse vectors, and interior points to dense vectors.}
%     \label{fig:sparsity}
% \end{figure}

\begin{figure*}[!t]
    \centering
    \includegraphics[width=.7\textwidth]{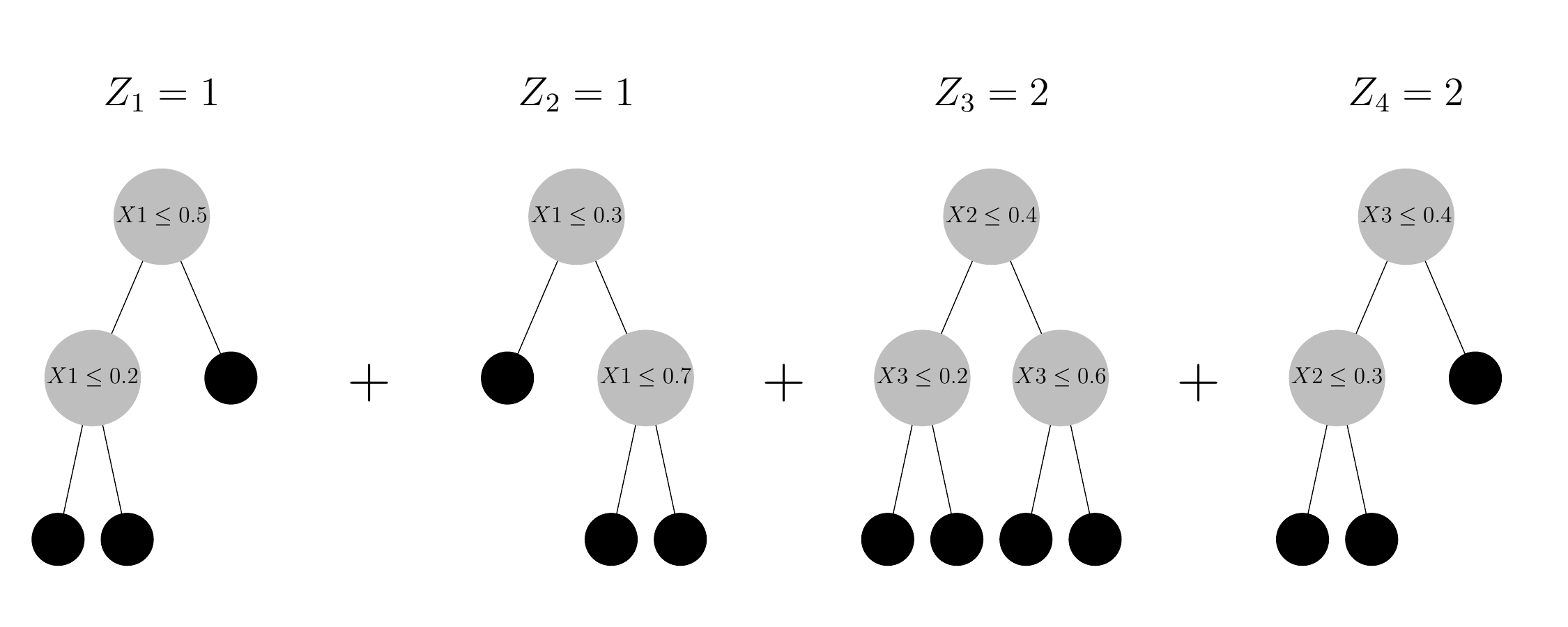}
    \caption{Schematic showing the effect of clustering trees within the ensemble. When $Z_t = 1$ split are constructed with $X_1$, but when $Z_t = 2$ splits are constructed with $(X_2, X_3)$.}
    \label{fig:illustration-1}
\end{figure*}

\citet{linero2017abayesian} also introduce the SBART model, which uses soft decision trees \citep{irsoy2012soft} which effectively replace the decision boundaries of BART with smooth sigmoid functions. This allows the SBART model to adapt to the smoothness level of $f(x)$; consequently, if $f_0(x)$ is assumed to be $\alpha$-H\"older, the posterior for the SBART model concentrates around $f_0(x)$ at close to the oracle minimax rate obtainable when the smoothness level is known a-priori. While the methodology we develop applies to the usual BART models, we will use the SBART model with the sparsity-inducing Dirichlet prior in all of our illustrations. 
% In the $P > N$ regime it is essential to use the sparsity-inducing Dirichlet prior, while the use of smooth decision trees is merely beneficial. Readers are referred to \citet{linero2017abayesian} for more details on the modifications to the BART prior we use in our illustrations; while these modificatins are not essential, they greatly improve the practical performance of BART on both simulated and real datasets. 

\section{DP-FORESTS}
\label{sec:dirichlet}

The distribution of $(\Tree_t, \sM_t)$ in the BART model is parameterized by the splitting proportions $s$, leaf variance $\sigma^2_\mu$, and tree topology parameters $(\gamma, \beta)$. To encourage a small number of low-order interactions, we specify a prior which clusters the trees into non-overlapping groups such that each cluster constructs splits using different subsets of the predictors. A schematic is given in Figure~\ref{fig:illustration-1} with $T = 4$. In this figure we see that the first two trees are dedicated to learning a main effect for $x_1$ while the second two trees are dedicated to learning an interaction between $x_2$ and $x_3$. 

We induce a clustering by using tree-specific splitting proportions $s\supp{t} \sim G$ and using a Dirichlet process prior on $G$ \citep{ferguson1973}. Specifically, we let $s\supp{t} \iid G$ conditional on $G$ and let $G \sim \DP(\omega G_0)$ where $G_0$ is a $\Dirichlet(\alpha w_1, \ldots, \alpha w_P)$ distribution and $\omega$ denotes the precision parameter of the Dirichlet process. Using the latent-cluster interpretation of the Dirichlet process (see, .e.g, \citealp{teh2006hierarchical}) this can be approximated by the following generative model:  

\begin{enumerate}
    \item Draw $\pi \sim \Dirichlet(\omega / K, \ldots, \omega / K)$ for large $K$.
    \item Draw $Z_1, \ldots, Z_T \indep \Categorical(\pi)$. 
    \item Draw $s\supp{1}, \ldots s\supp{K} \indep \Dirichlet(\alpha w_1, \ldots, \alpha w_P)$ where 
    $\sum_{p=1}^P w_p = 1, w_p \ge 0$.
    \item For $t = 1, \ldots, T$, draw $(\Tree_t, \sM_t)$ as described in Section~\ref{sec:review} with $s = s\supp{Z_t}$. 
\end{enumerate}

The $Z_t$'s cluster trees such that the trees within each group capture a single low-order interaction. Note that the use of the the sparsity inducing prior in step 3 above ensures that each $s\supp{k}$ will be nearly-sparse, and hence the trees with $Z_t = k$ will split on only a small subset of the predictors. The role played by this weight vector $w$ is to encourage a subset of the predictors to appear in multiple \emph{different} interactions. For example, if there are interactions $(X_1, X_2)$ and $(X_2, X_3)$ we do not want to encourage an additional $(X_1, X_3)$ interaction. A large value of $w_2$ allows for this by encouraging $X_2$ to appear in several interactions.

\subsection{Properties of the Prior}

The degree of sparsity within each cluster of trees, as well as the overall number of clusters used, are determined by the hyperparameters $\alpha$ and $\omega$. These hyperparameters are key in determining the interaction structures that the prior favors. 
To help anchor intuition we first consider several special cases of the DP-Forests model. First, we consider the behavior of the prior as $\alpha \to 0$ with $\omega$ fixed. In this case, with high probability each $s\supp{t}$ will have only one non-sparse entry. Consequently, each tree in the ensemble will split on at most one predictor. Because the trees are composed additively, this implies that none of the variables interact, and hence the prior concentrates on a sparse generalized additive model (SPAM, \citealp{ravikumar2007spam}). On the other hand, as $\alpha \to \infty$ we see that $s\supp{t} \to (w_1, \ldots, w_P)$ so that the prior reverts to original BART model with splitting proportions given by $(w_1, \ldots, w_P)$  described by \citet{bleich2014variable}. 

We can conduct a similar analysis with $\alpha$ fixed and $\omega$ with $K \to \infty$. As $\omega \to \infty$, each tree will be associated to a unique $s\supp{t}$. As $\omega \to 0$, on the other hand, all of the trees share the same $s\supp{t}$ so that the model collapses to the Dirichlet additive regression trees model described by \citet{linero2016bayesian}. 

% \subsection{The interaction structure prior}

The key difference between BART and a DP-Forest is that, once two variables are included, BART does not penalize interactions. Let $A_i$ and $A_j$ denote the event that variable $i$ and $j$ are included in the model, let $A_{ij}$ denote the event that variables $i$ and $j$ interact, and let $\Pi_{\alpha,\omega}$ denote the joint prior distribution for $\Tree_1, \ldots, \Tree_T$. We study the prior on the interaction structure by examining the probabilities 
\begin{math}
    \Lambda(\alpha, \omega) = \Pi_{\alpha, \omega}(A_{ij} \mid A_i \cap A_j), 
\end{math}
and
\begin{math}
  \Xi(\alpha,\omega) = \Pi_{\alpha,\omega}(A_{ik} \mid A_{ij} \cap A_{kj}).
\end{math}
In words, $\Lambda$ is the probability that $(i,j)$ interact given that both variables are relevant, while $\Xi$ represents the probability that $(i,k)$ interact given that $(i,j)$ and $(k,j)$ interact. Additionally, we examine the relationship between the average number of two-way interactions included in the model and the number of variables included.

\begin{figure}[!t]
    \centering
    \includegraphics[width=.45\textwidth]{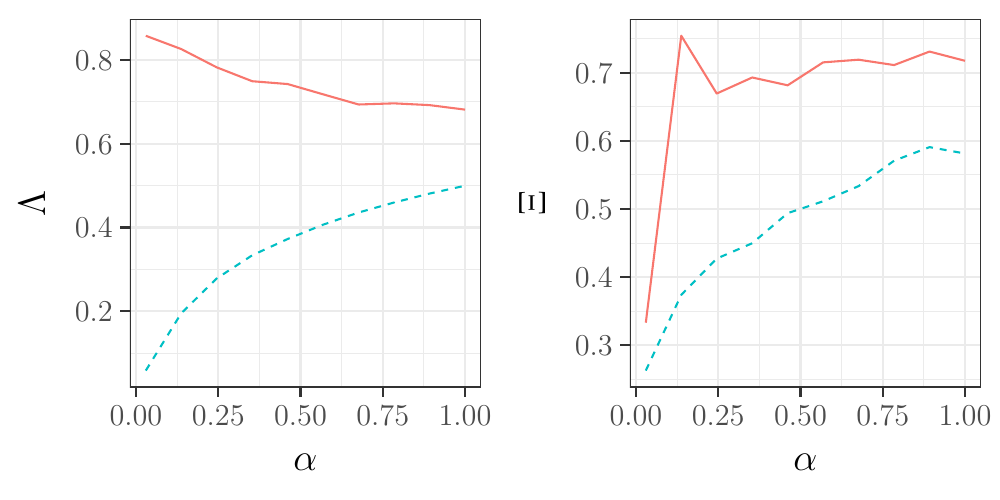}
    \caption{Plots of various quantities for $\omega = 0$ (solid, corresponding to SBART) and $\omega = 1$ (dashed) with $P = 5$ and $T = 50$. Left: plot of $\alpha$ against $\Lambda$. Middle: Plot of $\alpha$ against $\Xi$. Right: plot of the number of variables included in the model against the number of interactions. % Values are computed approximately by sampling from the prior distribution.
    }
    \label{fig:testing-prior}
\end{figure}

Figure~\ref{fig:testing-prior} shows several relationships between these quantities as $\alpha$ varies for both SBART and DP-Forests. We see that $\Lambda$ is quite large for all values of $\alpha$ with SBART, implying that the prior expects any variables included in the model to interact; the trend is decreasing in $\alpha$ only because a larger number of predictors will be included in the model, causing variables to compete for branches in the ensemble. DP-Forests do not encourage the inclusion of interactions, particularly when $\alpha$ is small. Next, we see that $\Xi$ is also uniformly large for SBART. This implies that the prior does not encourage interaction structures like the truth from Figure~\ref{fig:interaction-graph}, while a DP-Forest with a small choice of $\alpha$ does. 
% Finally, we see that for a given number of variables included in the model, the number of interactions is higher for SBART than for a DP-Forest. 

\subsection{Default Prior Settings}

A benefit of the BART framework is the existence of default priors which require minimal tuning from users. Where applicable, we do not stray from the defaults recommended in Section~\ref{sec:review}. Specific to DP-Forests, the key parameter controlling the behavior of the model is $\alpha$. On the basis of Figure~\ref{fig:testing-prior} we recommend choosing $\alpha$ to be small; we have found setting $\alpha \sim \Exponential$ with mean $0.1$ to work well. Conversely, in our illustrations the results for the DP-Forest model do not depend strongly on $\omega$, and we set $\omega \sim \Exponential(1)$. This leaves the weight vector $w = (w_1, \ldots, w_P)$ to be specified. In our illustrations, we first run a screening step which removes irrelevant predictors. In principle any method can be used for screening; in our illustrations, we use SBART to screen variables which have posterior inclusion probability below $50\%$, and set $w_j \propto I(\text{$j$ is not screened})$. A more principled alternative is to use another sparsity-inducing prior on $w$ but we do not pursue this strategy here. 

\subsection{Computation and Inference}
\label{sec:computation}

Inference for the DP-Forest model can be carried out using a Gibbs sampler with the Bayesian backfitting approach of \citet{chipman2010bart}. The Gibbs sampler operates on the state space $(\{\Tree_t, \sM_t, Z_t\}_{t = 1}^T, \{s\supp{k}, \pi_k\}_{k=1}^K, \alpha, \omega, \sigma^2_\mu, \sigma^2)$. We use standard Metropolis-within-Gibbs proposals to update $\Tree_t$ and $\sM_t$; see \citet{kapelner2014bartmachine} and \citet{pratola2016efficient} for details.
The parameters $\alpha$, $\omega$, $\sigma^2_\mu$, and $\sigma^2$ can all be updated easily using the slice sampling algorithm of \citet{slicesampling}. Finally, $Z_t, s\supp{k}$, and $\pi$ all have conjugate full-conditional distributions:  
% Metropolis-within-Gibbs is used to update the tree parameters $\Tree_t$ and $\sM_t$. We use (i) the ``birth'' and (ii) ``death'' Metropolis-Hastings proposals outlined by \citet{kapelner2014bartmachine} which incorporate the splitting proportion vector $s$ (see also \citet{chipman1998bayesian}). Additionally, we use (iii) the ``perturb'' proposal described by \citet{pratola2016efficient}. We refer to these works for precise details on the computation of the Metropolis-Hastings ratios, as well as \citet{linero2017abayesian} for the modifications required for the SBART algorithm. Finally,  we also (iv) propose draws of $\Tree_t$ from its prior given $s\supp{Z_t}$. 

\textbf{Full conditional for $\pi$:}
Note that $\pi$ is conditionally independent of all parameters given $(\omega, Z)$. By conjugacy of the Dirichlet distribution to multinomial sampling we have the full conditional $\pi \sim \Dirichlet(\omega/K + m_1, \ldots, \omega/K + m_K)$ where $m_k = \sum_t I(Z_t = k)$. 

\textbf{Full conditional for $s\supp{k}$:} 
The conjugacy of the Dirichlet prior to multinomial sampling implies a Dirichlet full-conditional when a single $s$ is used. To account for the clustering, we only consider the branches associated to trees with $Z_t = k$, giving the full conditional 
$s\supp{k} \sim \Dirichlet(\alpha w_1 + c_1\supp{k}, \ldots, \alpha w_P + c_P\supp{k})$ where $c_j\supp{k}$ is the number of branches associated to cluster $k$ which split on predictor $j$. 

\textbf{Full conditional for $Z_t$:}
Let $p(k)$ denote the full conditional for $Z_t$. The term $[Z_t = k]$ comes in only through the factors $\pi_k$ (the prior probability of $Z_t = k$) and $\prod_{j = 1}^P s_j^{(k)c_{tj}}$ where $c_{tj}$ is the number of branches of tree $t$ which split on predictor $j$ (the likelihood of tree $t$ having split on the predictors that it has, give $Z_t = k$). Hence $p(k) \propto \pi_k \prod_{j = 1}^P s_j^{(k)c_{tj}}$.

Putting these pieces together, we arrive at Algorithm~\ref{alg:bayesian-backfitting}, which describes a single iteration of the Gibbs sampler. 
\begin{algorithm}[h]
  \caption{Bayesian backfitting algorithm} \label{alg:bayesian-backfitting}
  \begin{algorithmic}[1]
    \For{$t = 1, \ldots, T$}
    \State Update $(\Tree_t, \sM_t)$ via Metropolis-Hastings.
    % \State Set \(Y_i^\star \gets Y_i - \sum_{k \ne t} g(X ; \Tree_{k} , \sM_k)\)
    % for \(i = 1, \ldots, N\).
    % \State Sample \(\Tree_t \sim \operatorname{Metrop}_{\Tree}(Y^\star, X, \tau_t, h)\). 
    % \State Sample \(\tau_t \sim \operatorname{Metrop}_{\tau}(Y^\star, X, \Tree_t, h)\).
    % \State Sample \(\sM_t \sim \Normal(\widehat \mu_t, \Omega_t)\) with
    % \((\widehat \mu_t, \Omega_t)\) described as in the supplementary material.
    \State Sample $Z_t \sim p(k), k = 1, \ldots, K$ where
    \begin{math}
      p(k) \propto \pi_k \prod_{j = 1}^P s_j^{(k)c_{tj}}
    \end{math}
    % \begin{align*}
    %   p(k) = \frac{\pi_k \prod_{j = 1}^P {s_j^{(k)c_{tj}}}}{\sum_{k'=1}^K \pi_{k'} \prod_{j=1}^P {s^{(k') c_{tj}}_j}}
    % \end{align*}
    and $c_{tj}$ is the number of branches associated to tree $t$ which split on predictor $j$.
    \EndFor
    \For{$k = 1,\ldots,K$}
    \State Sample \(s\supp{k} \sim \Dirichlet(\alpha w_1 + c\supp{k}_1, \ldots, \alpha w_P + c\supp{k}_P)\) where $c\supp{k}_j$ is the number of branches associated to cluster $k$ which split on predictor $j$. 
    \EndFor
    \State Sample $\pi \sim \Dirichlet(\omega / K + m_1, \ldots, \omega/K + m_K)$ where $m_k = \sum_{t = 1}^T I(Z_t = k)$.
    \State Sample $(\sigma, \sigma_{\mu}, \alpha, \omega)$ using slice sampling.
  \end{algorithmic}
\end{algorithm}

\section{EXPERIMENTS}

We now compare DP-Forests to existing methods on a number of synthetic datasets. We consider the following methods in addition to DP-Forests and SBART. 

\textbf{Additive groves}: The additive groves procedure of \citet{sorokina2008detecting}. Because tuning of the additive groves algorithm is compute-intensive, we ran several pilot studies to choose appropriate tuning parameters which perform well for the given simulation settings. % ($\alpha = 0.1, N = 3, B = 100$; see \citet{sorokina2008detecting}).  

\textbf{Hierarchical group lasso}: The hierarchical group lasso proposed by \citet{lim2015learning} for interaction detection;
% as implemented in the \texttt{glinternet} package on \texttt{CRAN}; 
we abbreviate this method by HL. This procedure was designed with linearity of $f_0(x)$ in mind. Tuning parameters are selected by cross-validation. 

\textbf{Hierarchical group lasso, least squares}: HL is used to \emph{select} the interactions and main effects, while the coefficients are estimated by least squares; we abbreviate this method by HL-LS. Tuning parameters are selected by cross validation. % (and may differ from the parameters of HL).

\textbf{Iterative random forests}: The iterative random forests (iRF) procedure proposed by \citet{basu2018iterative} as implemented in the \texttt{iRF} package on \texttt{CRAN}. We use the default $T = 500$ trees and 10 iterations of the iRF algorithm.

\begin{figure}[t]
    \centering
    \includegraphics[width=0.5\textwidth]{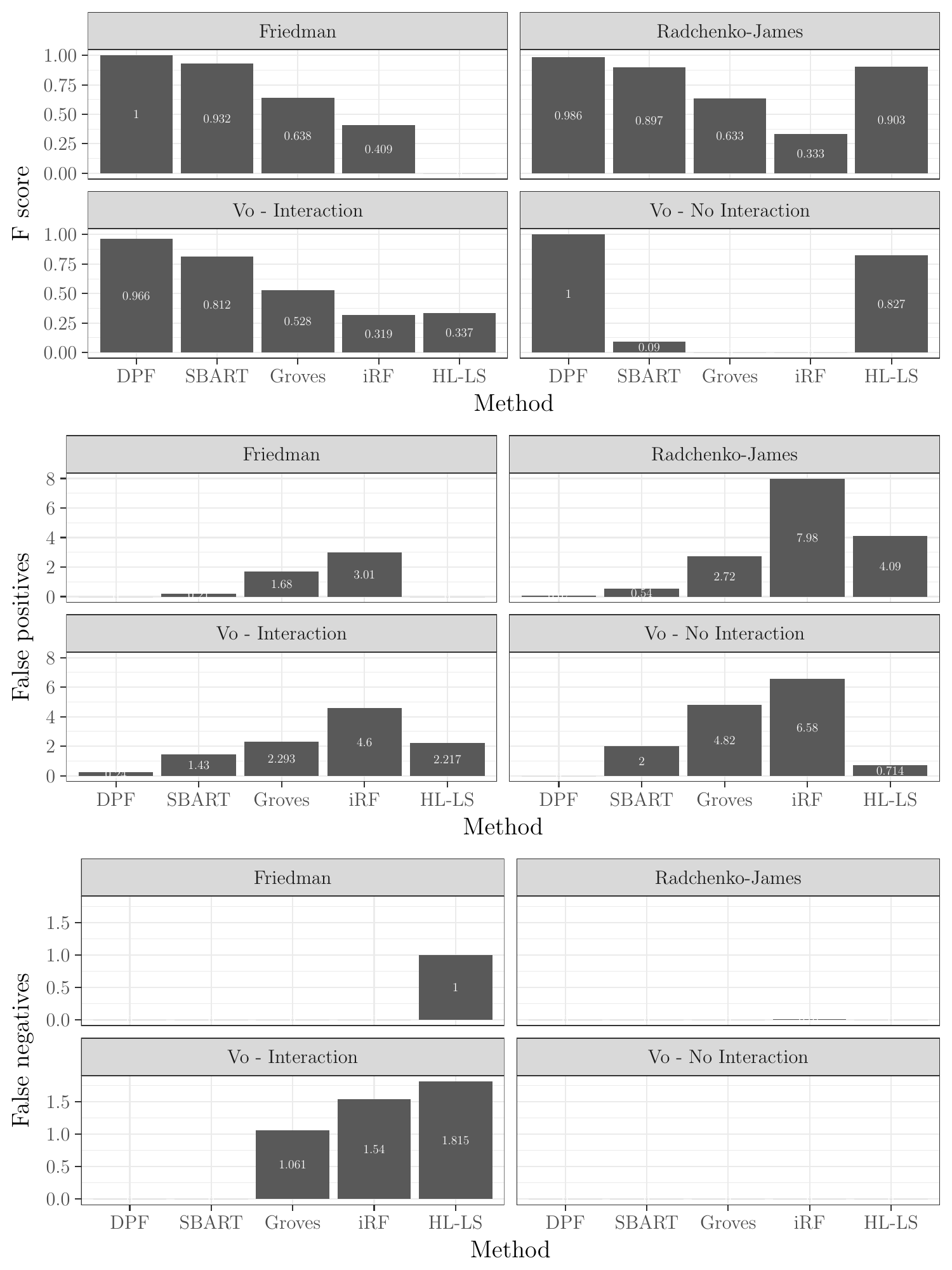}
    \caption{Barplot of results for interaction detection. The top row gives the average $F_1$ score for each method for detecting interactions. The second row gives the average number of false positive interactions detected. The bottom row gives the average number of false negatives detected. The average for each method is given on each bar.}
    \label{fig:sim_results_1_1}
\end{figure}
% Each of the above methods is designed to capture interactions in $f_0(x)$. We evaluate the methods according to the number of false positives and negatives for detecting both main effects and interaction terms. Additionally, we consider the root mean-squared error for each method 
% \begin{math}
%     \operatorname{RMSE} 
%     = 
%     \{ \int (f_0(x) - \widehat f(x))^2 \, F_0(dx)\}^{1/2}
% \end{math}
% where $X_i \sim F_0$ and $\widehat f(x)$ is an estimate of $f_0$.  

% Each simulation generates $X_i \in \Reals^P$ and then sets 
% \begin{align*}
%     Y_i = f_0(X_i) + \epsilon_i, \qquad \epsilon_i \sim \Normal(0, \sigma^2),
% \end{align*}
% with the functional form of $f_0(x)$, distribution of $X_i$, and value of $\sigma^2$ varying across each setting. 
Our simulation settings are borrowed from several existing works; we do not compare our methods to these other works due to a lack of publicly available software. 
\begin{enumerate}
    \item[(S1)] \citep{radchenko2010variable} We generate $X_i \sim \Uniform([0,1]^P)$ where $P = 50$, $N = 300$, and $\sigma^2 = 1$. We let $f_0(x)$ be
    % \begin{align*}
    %     f_0(x) = \sqrt{0.5} \bigg[\sum_{v = 1}^V f_v(x) + f_1(x) f_2(x) + f_1(x) f_3(x)\bigg]
    % \end{align*}
    \begin{align*}
        \sqrt{0.5} \bigg[\sum_{v = 1}^V f_v(x) + f_1(x) f_2(x) + f_1(x) f_3(x)\bigg]
    \end{align*}
    where $f_1(x) = x_1$, $f_2(x) = (1+x_2)^{-1}$, $f_3 = \sin(x_3)$, $f_4(x) = e^{x_4}$, and $f_5(x) = x_5^2$.
    Each $f_v(x)$ is further centered and scaled so that $E(f_v(X_i)) = 0$ and $\Var(f_v(X_i)) = 1$.
    \item[(S2)] \citep{vo2016sparse} We generate $X_i \sim \Normal(\zeros, \Identity)$ with $N = 100$, $P = 100$, and $\sigma = 0.14$. We let 
    \begin{math}
        f_0(x) = x_1 + x_2^2 + x_3 + x_4^2 + x_5 + x_1x_2 + x_2x_3 + x_3x_4. 
    \end{math}
    \item[(S3)] Same as (S2), but without the interaction effects. 
    \item[(S4)] \citep{friedman1991multivariate} A common test case for BART, we generate $X_i \sim \Uniform([0,1]^P)$ with $P = 250, N = 250$, and $\sigma^2 = 1$. We set
    \begin{math}
        f_0(x) = 10 \sin(x_1x_2) + 20(x_3 - 0.5)^2 + 10 x_4 + 5 x_5. 
    \end{math}
    % \begin{align}
    %     \label{eq:friedman}
    %     f_0(x) = 10 \sin(x_1x_2) + 20(x_3 - 0.5)^2 + 10 x_4 + 5 x_5. 
    % \end{align}
\end{enumerate}

\begin{figure}[t]
    \centering
    \includegraphics[width=0.5\textwidth]{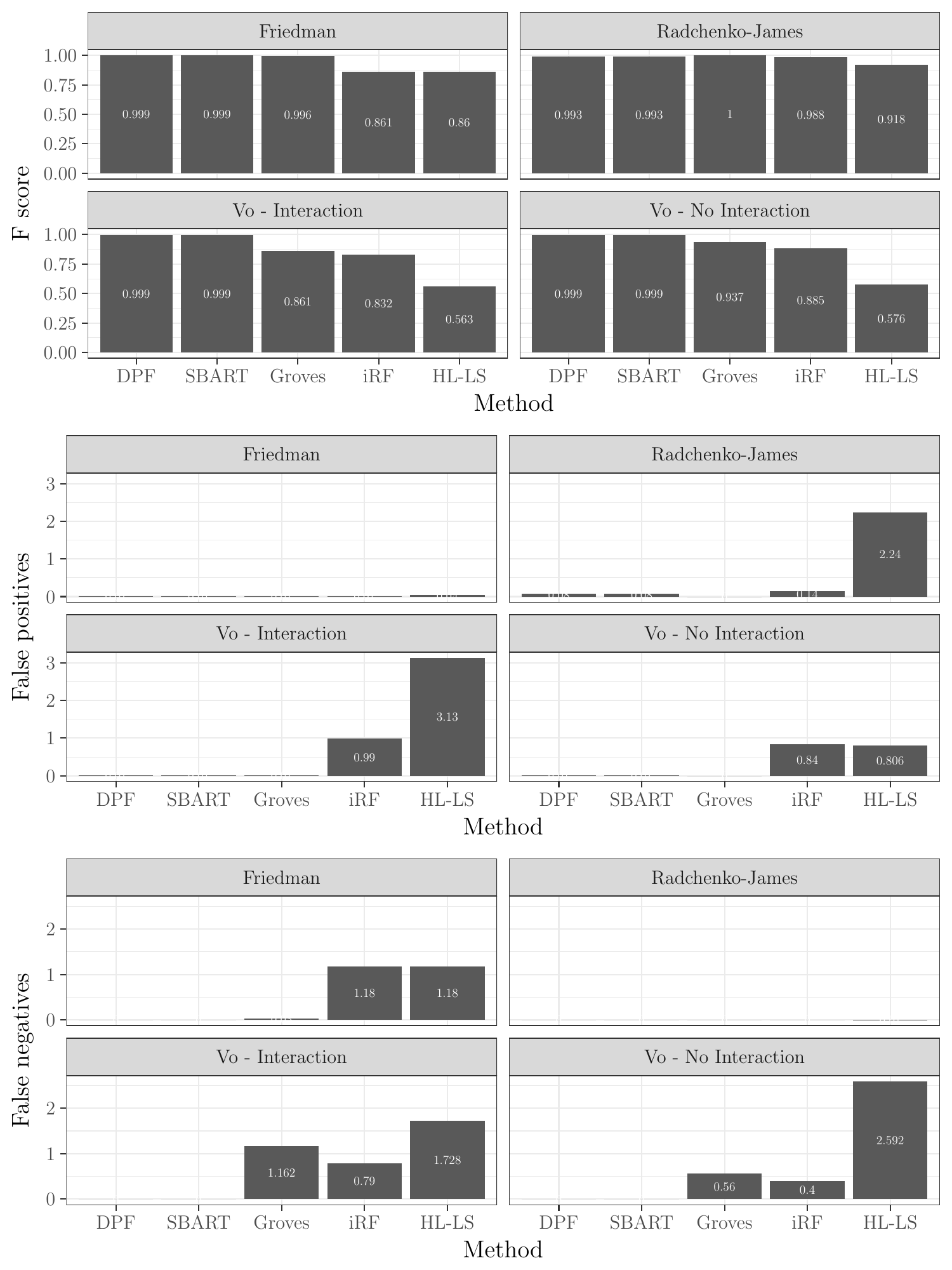}
    \caption{Barplot of results for detecting main effects.}
    \label{fig:sim_results_1_2}
\end{figure}

Each of these scenarios was replicated $100$ times. We evaluate each method according to the average number of false positives (FPs), false negatives (FNs), $F_1$ score, and integrated root-mean squared error $\|f_0 - \widehat f\|_2$. The $F_1$ score is a commonly used measure of overall accuracy that balances false positives against false negatives in variable selection tasks; see, for example, \citet{zhang2015cross}.

% Define the \emph{precision} and \emph{recall} by 
% $\Prec = \TP/(\TP + \FP)$ and $\Rec = \TP/(\TP + \FN)$
% % \begin{align*}
% %     \text{Prec} = \frac{\TP}{\TP + \FP}, 
% %     \qquad 
% %     \Rec = \frac{\TP}{\TP + \FN},
% % \end{align*}
% where $\TP, \FP, \FN$ denote the true positives, false positives, and false negatives respectively. The $F_1$ score is defined to be the harmonic mean of $\Prec$ and $\Rec$, i.e., $F_1 = 2 \, \Prec \, \Rec / (\Prec + \Rec)$. The $F_1$ score is a commonly used measure of overall accuracy that balances false positives against false negatives in variable selection tasks; see, for example, \citet{zhang2015cross}. We also evaluate each procedure according to their integrated root mean-squared error $\{\int (f_0 - \widehat f)^2 \, dF_0\}^{1/2}$ where $F_0$ is the distribution of $X_i$ and $\widehat f$ is an estimate of $f_0$.

Results for interaction detection are given in Figure~\ref{fig:sim_results_1_1}. We omit the results for HL because HL-LS performs uniformly better. Under all simulation settings, DP-Forests perform better than all other methods according to $F_1$ score. SBART is also competitive with other procedures on many of the datasets. As expected, the primary problem with SBART is that it has a relatively large number of false positives, i.e. it is susceptible to detecting spurious interactions. This issue is most pronounced on (S2) and (S3), with SBART detecting between 1.5 and 2 spurious interactions. % In terms of $F_1$ score, this is a disaster when there are no interactions present in the data, with SBART attaining an $F_1$ score of $0$ whenever it detects an interaction. 

Additive groves and iterative random forests generally perform worse than SBART. In addition to having a larger false positives rate, these procedures are also prone to false negatives under simulation (S2). With the exception of (S1), the hierarchical group-lasso (HL-LS) performs worse than the other methods. Under (S1), HL-LS has reasonable performance as each component of $f_0(x)$ can be reasonably well-approximated by the assumed linear model. HL-LS also appears to perform well under (S3); this, however, is due to the fact that HL-LS typically misses several main effects, which is a substantially worse outcome than detecting a spurious interaction. The nonlinearities under (S2) and (S4) also create problems for HL-LS.% ; in addition to missing the associated main effects, HL-LS misses the associated interactions.

All methods perform better for detecting the main effects. SBART and DP-Forests give identical results for the main effects due to the use of SBART in screening for DP-Forests. (S1) is the easiest setting, with all methods having very few false negatives and HL-LS the only method having non-negligible false-positives. 
% ; it is also the only setting where an alternative method (additives groves) outperforms DP-Forests and SBART
Under (S2), the non-Bayesian procedures all have non-negligible false negatives, and iRF and HL-LS are additionally prone to false positives; the story is similar under (S3), with HL-LS performing better in terms of false positives but worse in terms of false negatives. All methods perform well in terms of false positives under (S4), however iRF and HL-LS also suffer from many false negatives. 

\begin{figure}[t]
    \centering
    \includegraphics[width=.5\textwidth]{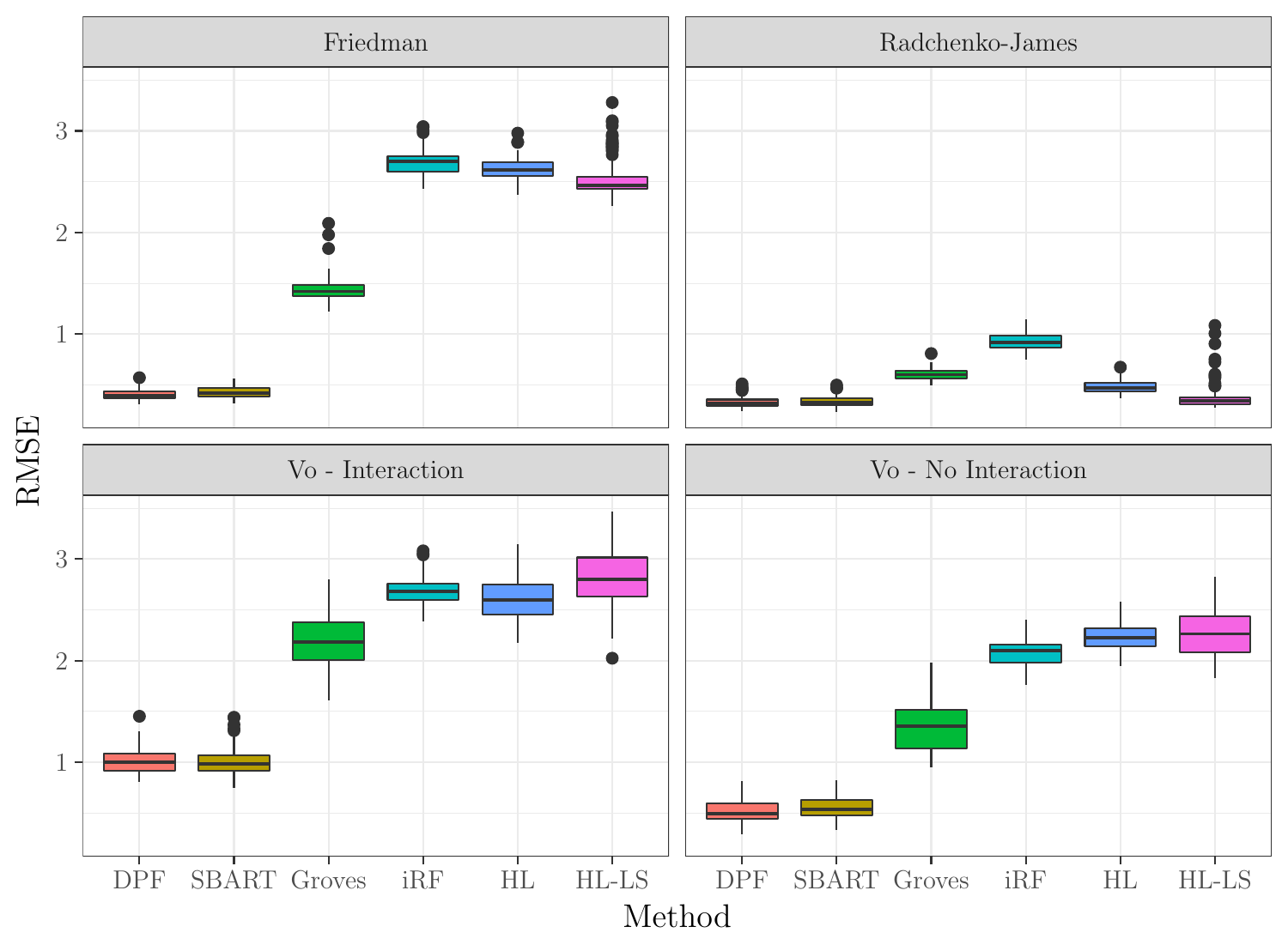}
    \caption{Boxplots given the distribution of integrated root mean-squared error for each method for each simulation setting.}
    \label{fig:sim_results_1_rmse}
\end{figure}

Results for assessing prediction performance in terms of integrated root mean-squared error (RMSE) are given in Figure~\ref{fig:sim_results_1_rmse}. SBART and DP-Forests perform very similarly in terms of RMSE. All other methods perform substantially worse under all settings. This is likely due to a multitude of factors. First, any false negatives will contribute to poor predictive performance. Second, SBART and DP-Forests are able to take advantage of underlying smoothness in the response function which additive groves and iterative random forests cannot, while HL and HL-LS suffer from an incorrect model specification. 
% Lastly, superior performance of SBART and DP-Forests is consistent with the general experience of BART-based methods outperforming boosting and random forests well on nonparametric regression tasks.

SBART and DP-Forests are competitive in terms of runtime. For example, on a single replicate of (S4), SBART and DP-Forests took 118 seconds and 241 seconds respectively to obtain 40,000 samples from the posterior. By comparison, iRF took 279 second, HL-LS took 91 seconds, and additive groves took 4966 seconds. Additive groves was by far the slowest procedure, due to the fact that recursive feature elimination is used. We conclude that, under these settings, DP-Forests outperform all competitors are a competitive computational budget.

We also consider the publicly available Boston housing dataset of \citet{harrison1978hedonic}. Analysis of the interaction structures present in this dataset was previously undertaken by \citet{radchenko2010variable} and \citet{vo2016sparse}. This dataset consists of $P = 13$ predictors and $N = 506$ neighborhoods, and a continuous response corresponding to the median house value in a given neighborhood.

\begin{table}
    \centering
    \begin{tabular}{lc}
    \toprule
    Method     & RMSE \\
    \midrule
    DP-Forests     &  1.00\\ 
    iRF     &  1.22\\ 
    HL      &  1.18\\ 
    Additive Groves & 1.16\\ 
    \bottomrule
    \end{tabular}
    \caption{Cross-validation estimate of root mean-squared prediction error on the Boston housing dataset normalized by the RMSE of the DP-Forest.}
    \label{tab:boston-rmspe}
\end{table}

We compare the methods in terms of goodness-of-fit, which is evaluated using a 5-fold cross validated estimate of root mean squared prediction error. Results are given in Table~\ref{tab:boston-rmspe}. For prediction, the DP-Forest and SBART outperform the competing methods. 

The DP-Forest includes most of the predictors in the model. This can be contrasted with the fit of a sparse additive model (SPAM) \citet{ravikumar2007spam} and the fit of the VANISH model reported by \citet{radchenko2010variable}, which include only a small number of predictors. Like the VANISH algorithm, the DP-Forest selects one interaction: there is strong evidence of an interaction between \texttt{DIS} (distance to an employment center in Boston) and \texttt{LSTAT} (the proportion of individuals in a neighborhood who are lower-status). This interaction was highly stable, and was selected by every fit to the data during cross-validation; additionally, this interaction was selected by additive groves in 4 out of 5 folds during cross-validation. Interestingly, this interaction was reportedly \emph{not} selected by VANISH, which instead selects an interaction between the variables \texttt{NOX} (nitrus-oxide concentration) and \texttt{LSTAT}. 

\begin{figure}[t]
    \centering
    \includegraphics[width = .5\textwidth]{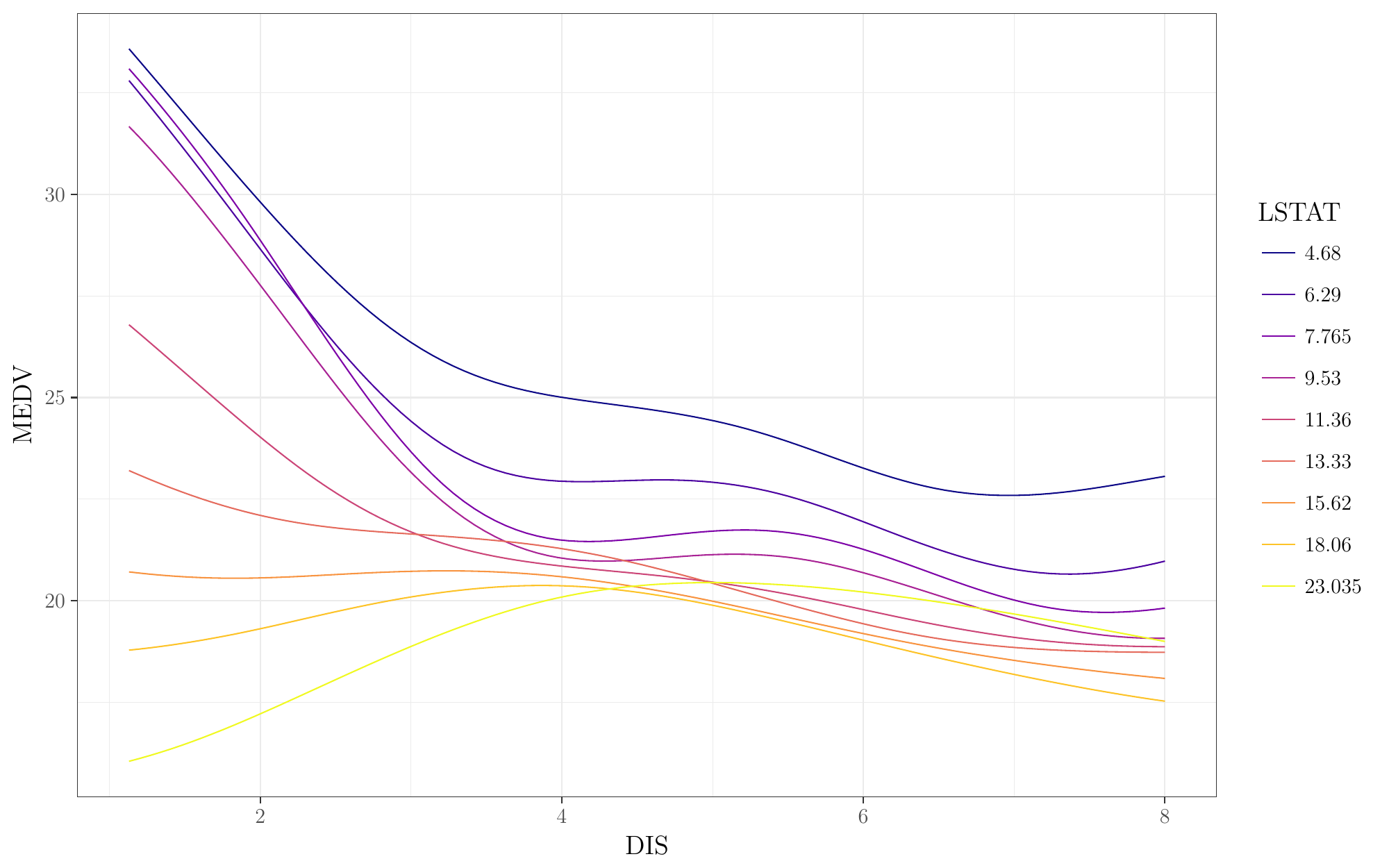}
    \caption{Graphical summary of the effect of distance \texttt{DIS} on \texttt{MEDV} for various values of \texttt{LSTAT}.}
    \label{fig:boston_results}
\end{figure}

Figure~\ref{fig:boston_results} gives a visualization of the \texttt{LSTAT}-\texttt{DIS} interaction. To summarize the interaction we use a ``fit-the-fit'' strategy and fit a generalized additive model to the fitted-values of the DP-Forest with a thin plate spline term for the interaction \citep{wood2003thin}. The plot then displays the \texttt{LSTAT}-specific effect of \texttt{DIS} for the $10^{\text{th}}, 20^{\text{th}}, \ldots, 90^{\text{th}}$ quantiles of \texttt{LSTAT}. This GAM nearly reproduces the fitted values from the DP-Forest and is easier to visualize. 
We see in Figure~\ref{fig:boston_results} a clear interaction between \texttt{DIS} and \texttt{LSTAT}. Intuitively, one expects that the closer a neighborhood is to an industry center the more expensive the housing will be. This is correct for areas with fewer lower-status individuals; however, this trend does not hold when there is a higher percentage of lower-status individuals. We remark also that the data is well supported near $0$ for all values of \texttt{LSTAT}, so that this behavior is unlikely to be due to extrapolation, though extrapolation may be an issue for large values of both \texttt{LSTAT} and \texttt{DIS}.

\section{DISCUSSION}
\label{sec:discussion}

We have introduced Dirichlet process forests (DP-Forests) and applied them to the problem of interaction detection. We demonstrated on both synthetic and real data that DP-Forests lead to improved interaction detection. Additionally, we demonstrated that DP-Forests are highly competitive with commonly used machine learning techniques for detecting low-order interactions.

There are a number of modifications one might make to improve performance further. One possibility is to allow $\sigma_{\mu}$ to also vary by mixture component. This would allow different mixture components to have different signal levels; for example, under simulation (S4), we would expect that a smaller value of $\sigma^2_{\mu}$ is appropriate for the mixture component responsible for $x_5$ relative to $x_4$. The proposed DP-Forests model captures this feature only indirectly through the number of trees assigned to each mixture component.

Additionally, it would be interesting to quantify the improvement in performance of DP-Forests over SBART theoretically. It is unknown whether SBART is variable-selection consistent, and establishing theoretically that DP-Forests are consistent for interaction detection while SBART is not remains an open problem.

% \subsubsection*{Acknowledgements}

% This work was partially supported by DOD grant 

\bibliography{mybib}

\begin{thebibliography}{}

\bibitem[Basu et~al., 2018]{basu2018iterative}
Basu, S., Kumbier, K., Brown, J.~B., and Yu, B. (2018).
\newblock iterative random forests to discover predictive and stable high-order
  interactions.
\newblock {\em Proceedings of the National Academy of Sciences}, page
  201711236.

\bibitem[Bien et~al., 2013]{bien2013lasso}
Bien, J., Taylor, J., and Tibshirani, R. (2013).
\newblock A lasso for hierarchical interactions.
\newblock {\em Annals of statistics}, 41(3):1111.

\bibitem[Bleich et~al., 2014]{bleich2014variable}
Bleich, J., Kapelner, A., George, E.~I., and Jensen, S.~T. (2014).
\newblock Variable selection for {BART}: An application to gene regulation.
\newblock {\em The Annals of Applied Statistics}, 8(3):1750--1781.

\bibitem[Chipman et~al., 2013]{chipman2013bayesian}
Chipman, H., George, E.~I., Gramacy, R.~B., and McCulloch, R. (2013).
\newblock Bayesian treed response surface models.
\newblock {\em Wiley Interdisciplinary Reviews: Data Mining and Knowledge
  Discovery}, 3(4):298--305.

\bibitem[Chipman et~al., 2010]{chipman2010bart}
Chipman, H.~A., George, E.~I., and McCulloch, R.~E. (2010).
\newblock {BART}: {B}ayesian additive regression trees.
\newblock {\em The Annals of Applied Statistics}, 4(1):266--298.

\bibitem[Ferguson, 1973]{ferguson1973}
Ferguson, T.~S. (1973).
\newblock A {B}ayesian analysis of some nonparametric problems.
\newblock {\em The Annals of Statistics}, 1:209--230.

\bibitem[Friedman, 1991]{friedman1991multivariate}
Friedman, J.~H. (1991).
\newblock Multivariate adaptive regression splines.
\newblock {\em The Annals of Statistics}, 19(1):1--67.

\bibitem[Hahn et~al., 2017]{hahn2017bayesian}
Hahn, P.~R., Murray, J.~S., and Carvalho, C. (2017).
\newblock Bayesian regression tree models for causal inference: regularization,
  confounding, and heterogeneous effects.
\newblock {\em arXiv preprint arXiv:1706.09523}.

\bibitem[Harrison and Rubinfeld, 1978]{harrison1978hedonic}
Harrison, D. and Rubinfeld, D.~L. (1978).
\newblock Hedonic prices and demand for clean air.
\newblock {\em Journal of Environmental Economics and Management}, 5:81--102.

\bibitem[Hastie, 2017]{hastie2017generalized}
Hastie, T.~J. (2017).
\newblock Generalized additive models.
\newblock In {\em Statistical models in S}, pages 249--307. Routledge.

\bibitem[Irsoy et~al., 2012]{irsoy2012soft}
Irsoy, O., Yildiz, O.~T., and Alpaydin, E. (2012).
\newblock Soft decision trees.
\newblock In {\em Proceedings of the International Conference on Pattern
  Recognition}.

\bibitem[Kapelner and Bleich, 2016]{kapelner2014bartmachine}
Kapelner, A. and Bleich, J. (2016).
\newblock {bartMachine}: Machine learning with {B}ayesian additive regression
  trees.
\newblock {\em Journal of Statistical Software}, 70(4):1--40.

\bibitem[Lim and Hastie, 2015]{lim2015learning}
Lim, M. and Hastie, T. (2015).
\newblock Learning interactions via hierarchical group-lasso regularization.
\newblock {\em Journal of Computational and Graphical Statistics},
  24(3):627--654.

\bibitem[Linero, 2016]{linero2016bayesian}
Linero, A.~R. (2016).
\newblock Bayesian regression trees for high dimensional prediction and
  variable selection.
\newblock {\em Journal of the American Statistical Association}.
\newblock To appear.

\bibitem[Linero, 2017]{linero2017review}
Linero, A.~R. (2017).
\newblock A review of tree-based {B}ayesian methods.
\newblock {\em Communications for Statistical Applications and Methods},
  24(6):543--559.

\bibitem[Linero and Yang, 2017]{linero2017abayesian}
Linero, A.~R. and Yang, Y. (2017).
\newblock Bayesian regression tree ensembles that adapt to smoothness and
  sparsity.
\newblock {\em arXiv preprint arXiv:1707.09461}.

\bibitem[Murray, 2017]{murray2017log}
Murray, J.~S. (2017).
\newblock Log-linear {B}ayesian additive regression trees for categorical and
  count responses.
\newblock {\em arXiv preprint arXiv:1701.01503}.

\bibitem[Neal, 2003]{slicesampling}
Neal, R.~M. (2003).
\newblock Slice sampling.
\newblock {\em The Annals of Statistics}, 31:705--767.

\bibitem[Pratola, 2016]{pratola2016efficient}
Pratola, M. (2016).
\newblock Efficient {M}etropolis-{H}astings proposal mechanisms for {B}ayesian
  regression tree models.
\newblock {\em Bayesian Analysis}, 11(3):885--911.

\bibitem[Radchenko and James, 2010]{radchenko2010variable}
Radchenko, P. and James, G.~M. (2010).
\newblock Variable selection using adaptive nonlinear interaction structures in
  high dimensions.
\newblock {\em Journal of the American Statistical Association},
  105(492):1541--1553.

\bibitem[Ravikumar et~al., 2007]{ravikumar2007spam}
Ravikumar, P., Liu, H., Lafferty, J., and Wasserman, L. (2007).
\newblock {SPAM}: Sparse additive models.
\newblock In {\em Proceedings of the 20th International Conference on Neural
  Information Processing Systems}, pages 1201--1208.

\bibitem[{Rockova} and {van der Pas}, 2017]{rockova2017posterior}
{Rockova}, V. and {van der Pas}, S. (2017).
\newblock Posterior concentration for {B}ayesian regression trees and their
  ensembles.
\newblock {\em {arXiv preprint arXiv:1078.08734}}.

\bibitem[Sorokina et~al., 2008]{sorokina2008detecting}
Sorokina, D., Caruana, R., Riedewald, M., and Fink, D. (2008).
\newblock Detecting statistical interactions with additive groves of trees.
\newblock In {\em Proceedings of the 25th international conference on Machine
  learning}, pages 1000--1007. ACM.

\bibitem[Sparapani et~al., 2016]{sparapani2016nonparametric}
Sparapani, R.~A., Logan, B.~R., McCulloch, R.~E., and Laud, P.~W. (2016).
\newblock Nonparametric survival analysis using {B}ayesian additive regression
  trees ({BART}).
\newblock {\em Statistics in medicine}.

\bibitem[Starling et~al., 2018]{starling2018functional}
Starling, J.~E., Murray, J.~S., Carvalho, C.~M., Bukowski, R., and Scott, J.~G.
  (2018).
\newblock Functional response regression with funbart: an analysis of
  patient-specific stillbirth risk.
\newblock {\em arXiv preprint arXiv:1805.07656}.

\bibitem[Teh et~al., 2006]{teh2006hierarchical}
Teh, Y.~W., Jordan, M.~I., Beal, M.~J., and Blei, D.~M. (2006).
\newblock Hierarchical {D}irichlet processes.
\newblock {\em Journal of the American Statistical Association},
  101(476):1566--1581.

\bibitem[Vo and Pati, 2016]{vo2016sparse}
Vo, G. and Pati, D. (2016).
\newblock Sparse additive {G}aussian process with soft interactions.
\newblock {\em arXiv preprint arXiv:1607.02670}.

\bibitem[Wang et~al., 2007]{wang2007tuning}
Wang, H., Li, R., and Tsai, C.-L. (2007).
\newblock Tuning parameter selectors for the smoothly clipped absolute
  deviation method.
\newblock {\em Biometrika}, 94(3):553--568.

\bibitem[Wood, 2003]{wood2003thin}
Wood, S.~N. (2003).
\newblock Thin plate regression splines.
\newblock {\em Journal of the Royal Statistical Society: Series B (Statistical
  Methodology)}, 65(1):95--114.

\bibitem[Zhang and Yang, 2015]{zhang2015cross}
Zhang, Y. and Yang, Y. (2015).
\newblock Cross-validation for selecting a model selection procedure.
\newblock {\em Journal of Econometrics}, 187(1):95--112.

\end{thebibliography}

\end{document}